\documentclass{emulateapj}

\usepackage{graphicx}
\usepackage{dcolumn}
\usepackage{bm}
\usepackage{amsfonts}
\usepackage{color}
\usepackage{amsmath}

\begin{document}

\shorttitle{A NEW GRAVITATIONAL-WAVE SIGNATURE OF SASI ACTIVITIES}
\shortauthors{Kuroda, Kotake \& Takiwaki}

\title{
A NEW GRAVITATIONAL-WAVE SIGNATURE FROM STANDING ACCRETION SHOCK INSTABILITIES IN SUPERNOVAE
}

\author{Takami Kuroda\altaffilmark{1}, Kei Kotake\altaffilmark{2,3} and Tomoya Takiwaki\altaffilmark{3}
}

\affil{$^1$Department of Physics, University of Basel,
Klingelbergstrasse 82, 4056 Basel, Switzerland}
\affil{$^2$Department of Applied Physics, Fukuoka University,
 8-19-1, Jonan, Nanakuma, Fukuoka, 814-0180, Japan}
\affil{$^3$Division of Theoretical Astronomy, National Astronomical Observatory of Japan, 2-21-1, Osawa, Mitaka, Tokyo, 181-8588, Japan}

\begin{abstract}
We present results from fully relativistic three-dimensional 
 core-collapse supernova (CCSN) simulations of a non-rotating
 $15 M_{\odot}$ star using three different nuclear equations of state 
(EoSs). From our simulations covering up to $\sim350$ ms after bounce,
we show that the development of the standing accretion 
shock instability (SASI) differs significantly depending 
on the stiffness of nuclear EoS. Generally, the SASI activity
  occurs more vigorously in models with softer EoS. 
By evaluating the gravitational-wave (GW) emission, 
we find a new GW signature on top of the previously
 identified one, in which the typical GW frequency 
increases with time due to an accumulating accretion 
 to the proto-neutron star (PNS).
 The newly observed quasi-periodic signal appears in the frequency range
 from $\sim 100$ to $200$ Hz and persists for $\sim 150$ ms 
before neutrino-driven convection dominates over the SASI. 
By analyzing the cycle frequency of the SASI sloshing
 and spiral modes as well as the mass accretion rate to the emission region, 
we show that the SASI frequency is correlated
 with the GW frequency. This is because the 
 SASI-induced temporary perturbed mass accretion strike
 the PNS surface, leading to the quasi-periodic GW 
emission. Our results show that the GW signal, which
 could be a smoking-gun signature of the SASI,
 is within the detection limits of LIGO, advanced 
Virgo, and KAGRA for Galactic events.
 \end{abstract}


\section{Introduction}
\label{sec:Introduction}
 Clarifying a correspondence between core-collapse 
supernova (CCSN) dynamics and the gravitational wave 
(GW) signals is a time-honored attempt since the 1980s
 \citep{EMuller82}. Very recently 
the observational horizon of GW astronomy extends
 far enough to allow the first detection coined
 by LIGO for the black hole merger event
 \citep{gw2016}.
Extensive research over the decades
has strengthened our confidence 
 that CCSNe, next to compact binary mergers, could also be 
 one of the most promising astrophysical sources 
of GWs 
(see \cite{Ott09,Kotake13} for reviews).

Traditionally most of the theoretical predictions have 
  focused on the GW signals 
from rotational core collapse and bounce (see, e.g., \citet{Dimmelmeier02B,Scheidegger10,Ott12,KurodaT14,Yokozawa15}).
In the post bounce phase, a variety of GW emission 
processes have been proposed,
 including convection inside 
 the proto-neutron star (PNS) and in the postshock region
 \citep{burohey}, 
the Standing-Accretion-Shock-Instability (SASI, \citet{Marek09,Kotake07,Kotake09,Murphy09}), and 
nonaxisymmetric instabilities \citep{Ott05,Scheidegger10,KurodaT14}.

In the non-rotating core,  \citet{Murphy09} firstly 
 showed in their two-dimensional (2D) models 
 that the evolution of convective activities 
in the PNS surface regions can be 
imprinted in the GW spectrogram. 
The characteristic GW frequency is considered 
as a result of the $g$-mode oscillation excited
 by the downflows to the PNS 
\citep{marek09gw} and by the deceleration of convection 
plumes hitting the surface \citep{Murphy09}.
 These features have also been identified in
 more recent 2D models 
 with best available neutrino transport scheme
 \citep{Yakunin10,BMuller13,Yakunin15}.
Furthermore \cite{BMuller13} showed
 in their self-consistent 2D models that the 
SASI motions become generally more violent for 
 more massive progenitors, which tends to make
 the GW amplitudes and frequencies higher. 

Not to mention the explosion dynamics (e.g., 
\citet{janka16,Takiwaki14,couch13a,Hanke12}),
 the GW signatures are very sensitive to the spatial 
dimension employed in the numerical modeling
 (e.g., \citet{Kotake09,EMuller12}).
  Due to the high numerical cost, however, only a few
 full three-dimensional 
(3D) models have been reported so far to study
 the postbounce GW features \citep[without 
any symmetry constraints and excision of the PNS, e.g.,]
[]{Scheidegger10,Ott12,KurodaT14}.
Using a prescribed boundary condition of the PNS contraction, \cite{Hanke13} showed in their 
3D models that a rapid shrinking of the PNS fosters the 
development of the SASI. General relativity (GR)
 should play a crucial role because the SASI is favored 
by smaller shock radii due to the short SASI's growth rate
 \citep{Foglizzo06}. To have a final word
 on recent hot debates about the impacts of  
neutrino-driven convection vs. the SASI on the 
 supernova mechanism (e.g., \citet{burrows13}),
 full 3D-GR models are needed, which is also the case 
for clarifying the GW emission processes.

In this {\it Letter}, we 
study the GW emission from a non-rotating 
$15 M_{\odot}$ star by performing 
 3D-GR hydrodynamic simulations with an approximate 
neutrino transport. Using three modern nuclear EoSs,
we investigate its impacts on both the postbounce dynamics 
and the GW emission. Our results reveal a new GW signature
 where the SASI activity is imprinted.
 We discuss the detectability of the signals, if detected,
 could provide the {\it live broadcast}
 that pictures how the supernova shock is dancing 
 in the core.

\begin{figure*}[htbp]
\begin{center}
\includegraphics[width=70mm,angle=-90.]{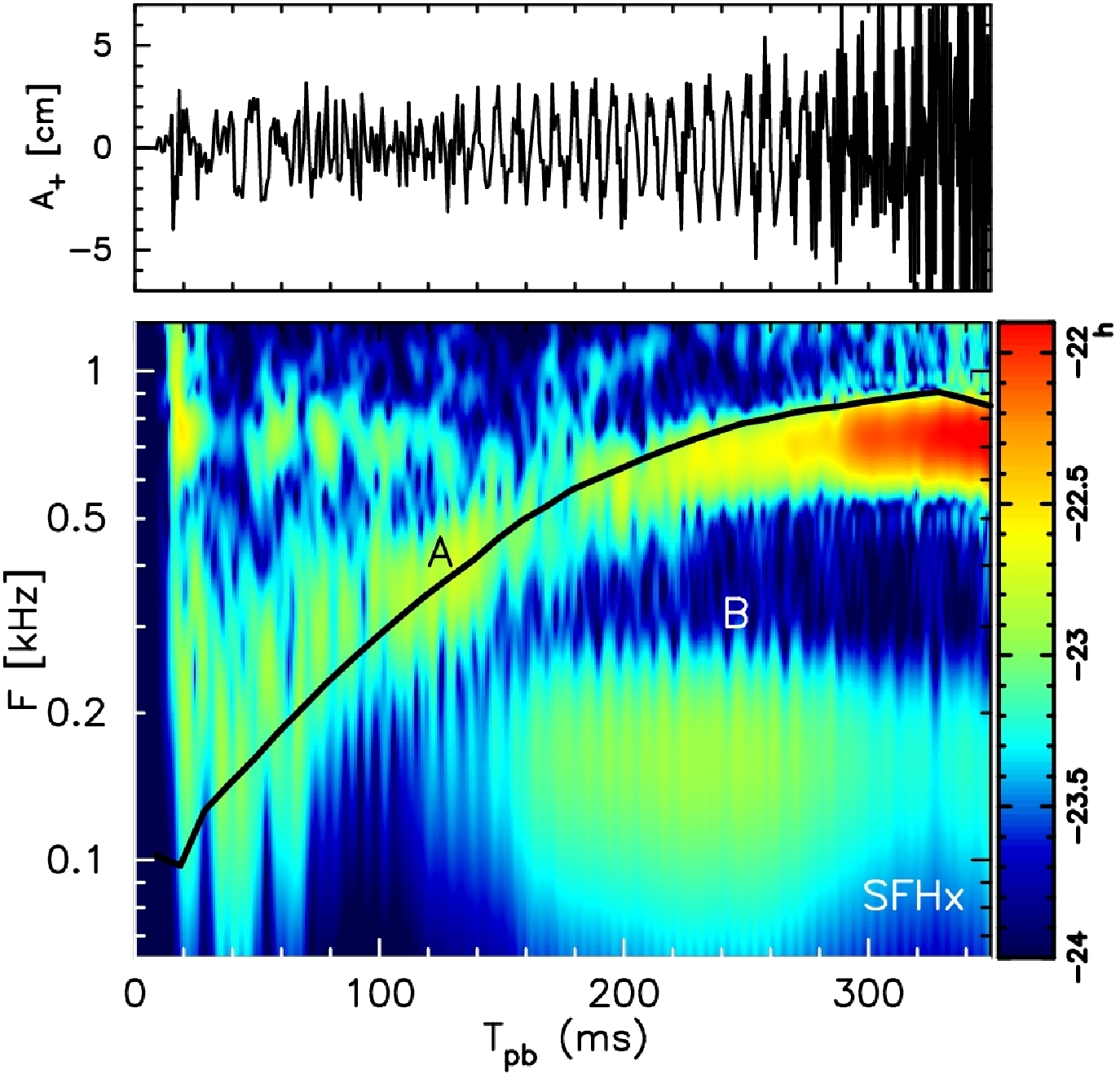}
\includegraphics[width=70mm,angle=-90.]{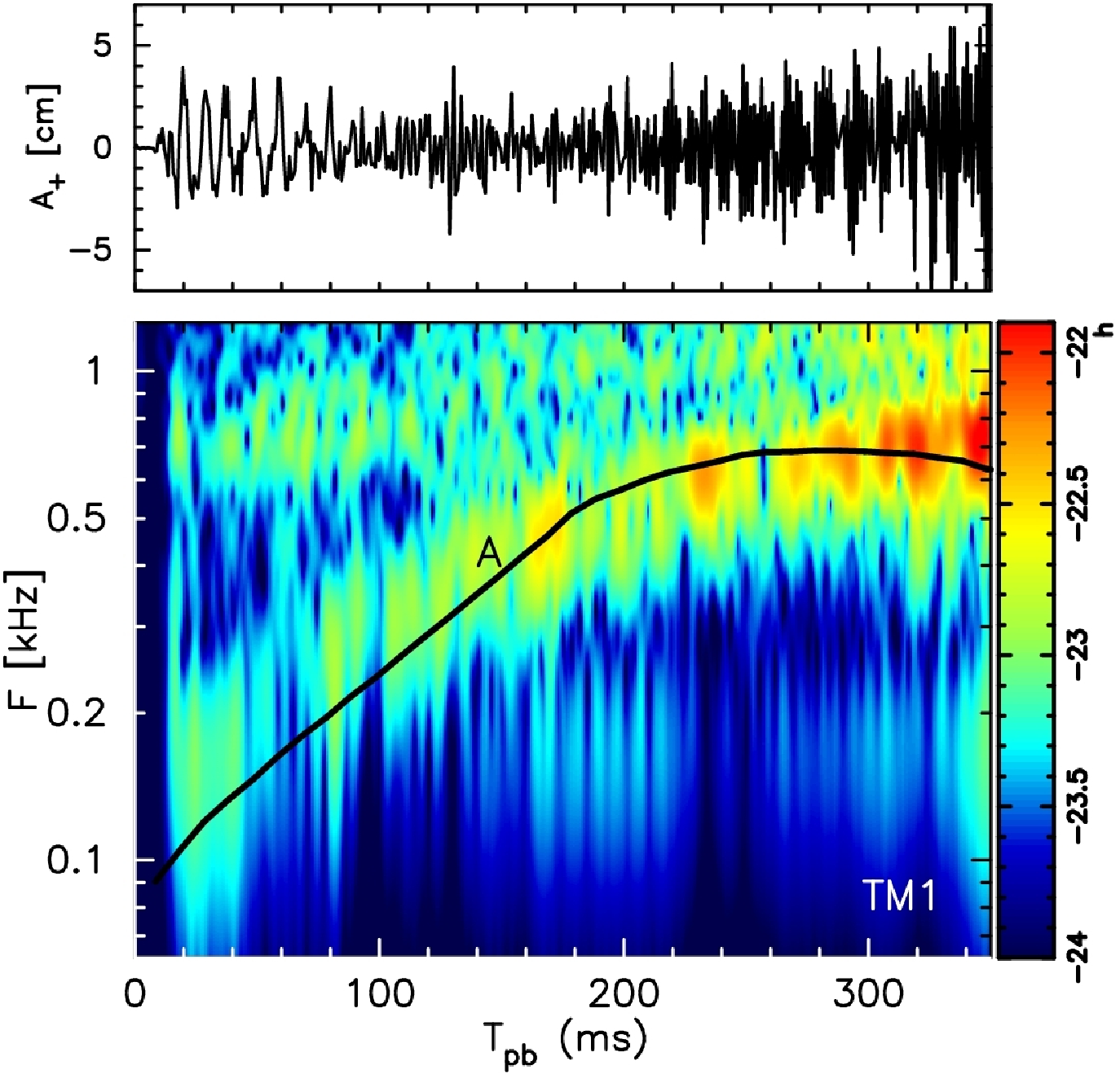}
  \caption{In each set of panels, we plot, top; gravitational wave amplitude of plus mode $A_+$ [cm],
  bottom; the characteristic wave strain in frequency-time domain $\tilde h$ in a logarithmic scale
  which is over plotted by the expected peak frequency $F_{\rm peak}$ (black line denoted by ``A'').
  ``B'' indicates the low frequency component.
  The component ``A'' is originated from the PNS $g$-mode oscillation \citep{Marek09,BMuller13}.
  The component ``B'' is considered to be associated with the SASI activities (see Sec. \ref{sec:Results}).
  Left and right panels are for TM1 and SFHx, respectively.
   We mention that SFHx (left) and TM1 (right) are softer and stiffer EoS models, respectively.
\label{GWs}
}
\end{center}
\end{figure*}
\section{Numerical Methods}
\label{sec:Numerical Methods}
In our full GR radiation-hydrodynamics simulations, we solve
the evolution equations of metric, hydrodynamics, 
and neutrino radiation.
Each of them is solved in an operator-splitting manner, but the system evolves self-consistently
as a whole satisfying the Hamiltonian and momentum constraints \citep{KurodaT12,KurodaT14}.

Regarding the metric evolution, we evolve the standard BSSN variables $\tilde\gamma_{ij}$, $\phi$, $\tilde A_{ij}$, $K$ and
$\tilde\Gamma^{i}$ \citep{Shibata95,Baumgarte99}.
The gauge is specified by the ``1+log'' lapse and by the Gamma-driver-shift condition.

In the radiation-hydrodynamic part, the total stress-energy tensor $T^{\alpha\beta}_{\rm (total)}$ is expressed as
\begin{equation}
T_{\rm (total)}^{\alpha\beta} = T_{\rm (fluid)}^{\alpha\beta} + \sum_{\nu\in\nu_e,\bar\nu_e,\nu_x}T_{(\nu)}^{\alpha\beta},
\label{TotalSETensor}
\end{equation}
where $T_{\rm (fluid)}^{\alpha\beta}$ and $T_{(\nu)}^{\alpha\beta}$ are the 
stress-energy tensor of fluid and neutrino radiation field, respectively.
All radiation and hydrodynamical variables are evolved in conservative ways.
We consider all three flavors of neutrinos ($\nu_e,\bar\nu_e,\nu_x$) with $\nu_x$ 
representing heavy-lepton neutrinos (i.e. $\nu_{\mu}, \nu_{\tau}$ and their anti-particles).
To follow the 3D hydrodynamics up to $\lesssim 400$ ms 
postbounce, we shall omit the energy dependence of the 
 radiation in this work (see, however, \cite{KurodaT16}).

We use three EoSs based on the 
relativistic-mean-field theory with different
 nuclear interaction treatments, which are DD2 and TM1 of \cite{HS} 
and SFHx of \cite{SFH}.

For SFHx, DD2, and TM1\footnote{The symmetry energy $S$ at nuclear saturation density
is $S=28.67$, 31.67, and 36.95 MeV, respectively. \citep[e.g.,][]{Fischer14}},
the maximum gravitational mass $M_{\rm max}$ and
the radius of cold NS $R$ in the vertical part of the mass-radius relationship are
 $M_{\rm max}=2.13$, 2.42, and, 2.21 $M_\odot$
and $R\sim12$, 13, and, 14.5 km, respectively \citep{Fischer14}.
SFHx is thus softest followed in order by DD2, and TM1.
Among these threes, while DD2 is consistent with nuclear experiments, such as, for its symmetry energy \citep{Lattimer13},
SFHx is the best fit model with the observational mass-radius relationship.
All EoSs are compatible with NS mass measurement $\sim2.04$ $M_\odot$ \citep{Demorest10}.
Our 3D-GR models are named as DD2, TM1 and SFHx, which simply reflects the EoS used.

We study a frequently used 15 $M_{\odot}$ star 
of \cite{WW95}.
The 3D computational domain is a cubic box with 15000 km width
and nested boxes with 8 refinement levels are embedded.
Each box contains $128^3$ cells and the minimum grid size near the origin is $\Delta x=458$m.
In the vicinity of the stalled shock front $R\sim100$ km, our resolution 
achieves $\Delta x\sim 1.9$ km, i.e., the effective angular resolution becomes $\sim1^\circ$.

Extraction of GWs from our simulations is done by the conventional quadrupole formula
in which the transverse and the trace-free gravitational field $h_{ij}$ is expressed by \citep{Misner73}
\begin{eqnarray}
\label{eq:hij}
h_{ij}(\theta,\phi)=\frac{A_+(\theta,\phi)e_++A_\times(\theta,\phi) e_\times}{D}
\end{eqnarray}
In Eq.(\ref{eq:hij}), $A_{+/\times}(\theta,\phi)$ represent amplitude of orthogonally polarized wave components
with emission angle $(\theta,\phi)$ dependence \citep{Scheidegger10,KurodaT14},
$e_{+/\times}$ denote unit polarization tensors and $D$ is the source distance where we set $D=10$ kpc in this paper.

\begin{figure*}[htpb]
\begin{center}
\includegraphics[width=100mm,angle=0.]{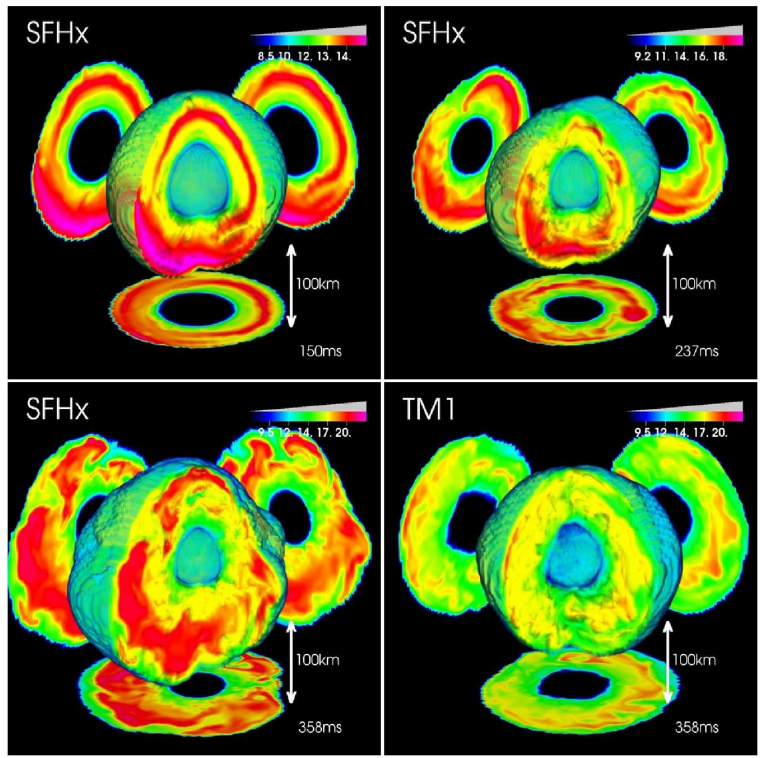}
\end{center}
  \caption{Snapshots of the entropy distribution ($k_{\rm B}$ baryon$^{-1}$) for models SFHx and TM1
(top left; $T_{\rm pb}=150$ ms of SFHx, top right; $T_{\rm pb}=237$ ms of SFHx, bottom left; $T_{\rm pb}=358$ ms  of SFHx, bottom right; $T_{\rm pb}=358$ ms of TM1).
The contours on the cross sections in the $x$ = 0 (back right), $y$ = 0 (back left), and $z$ = 0 (bottom) planes are, respectively projected on the sidewalls of the graphs.
The 90$^\circ$ wedge on the near side is excised to see the internal structure.
Note that to see the entropy structure clearly in each dynamical phase, we change the maximum entropy
in the colour bar as $s_{\rm max}=16$, 20 and 22 $k_{\rm B}$ baryon$^{-1}$ for $T_{\rm pb}=150$, 237 and 358 ms, respectively.}
\label{3Dimages}
\end{figure*}

\section{Results}
\label{sec:Results}
We start by describing the hydrodynamics at bounce.
The central rest mass density $\rho_c$ 
reaches $\rho_c=3.69,$ 3.75 and 4.50 $\times10^{14}$ g cm$^{-3}$
for TM1, DD2 and SFHx, which is higher as expected
 for the softer EOS  (e.g., \citet{Fischer14}).

After bounce, the non-spherical matter motion develops and starts GW emission.
In Fig. \ref{GWs}, we plot time evolution of the angle dependent
GW amplitude (only plus mode $A_+(\theta,\phi)$, black line) in upper panel and the characteristic wave strain
in frequency-time domain $\tilde h(\theta,\phi,F)$\citep[see Eq.(44) in][]{KurodaT14} in lower one.
Here $F$ denotes the GW frequency.
We extract GWs along the north pole $(\theta,\phi)=(0,0)$.
The post bounce hydrodynamics evolutions in DD2 are rather similar to TM1 and
we mainly focus on the comparison between SFHx and TM1 in the following.

The GW amplitude ($A_+$, upper two-panels) shows a consistent behavior as reported in \cite{BMuller13,Ott13,Yakunin15}.
It shows initial low frequency and slightly larger amplitude till $T_{\rm pb}\sim60$ ms,
which is followed by a quiescent phase with higher frequency till $T_{\rm pb}\sim150$ ms.
Afterward the amplitude and frequency become larger with time.

From spectrograms (lower panels), we see a narrow band spectrum (labeled as ``A'' in both models)
which shows an increasing trend in its peak frequency.
\cite{BMuller13,Murphy09} showed that this peak shift can be explained by properties of PNS, such as its compactness and surface temperature.
By following Eq.(17) in \cite{BMuller13}, we overplot $F_{\rm peak}$ in lower panels (black line).
In both models $F_{\rm peak}$ indeed tracks spectral peak quite well,
although there is some exception in late phase of SFHx ($T_{\rm pb}\ga200$ ms)
when the other strong component appears at $100\la F\la200$ Hz (labeled as ``B'').
The component ``A'' is thus actually originated from the $g$-mode oscillation of the PNS surface.


Before going into detail to explain the origin of the low frequency 
component ``B'', we briefly focus on several key differences in the 
hydrodynamcs evolution between SHFx and TM1.
 In Fig.\ref{3Dimages}, SFHx experiences violent sloshing (top-left)
and spiral motions of the SASI (top-right), before neutrino-driven
convection dominates over the SASI (bottom-left), whereas 
the SASI activities are less developed in TM1.
For SFHx, the clear SASI motions are observed after 
the prompt convection phase ceases at $T_{\rm pb} \sim 50$ ms.

\begin{figure}[htpb]
\begin{center}
\includegraphics[width=100mm,angle=-90.]{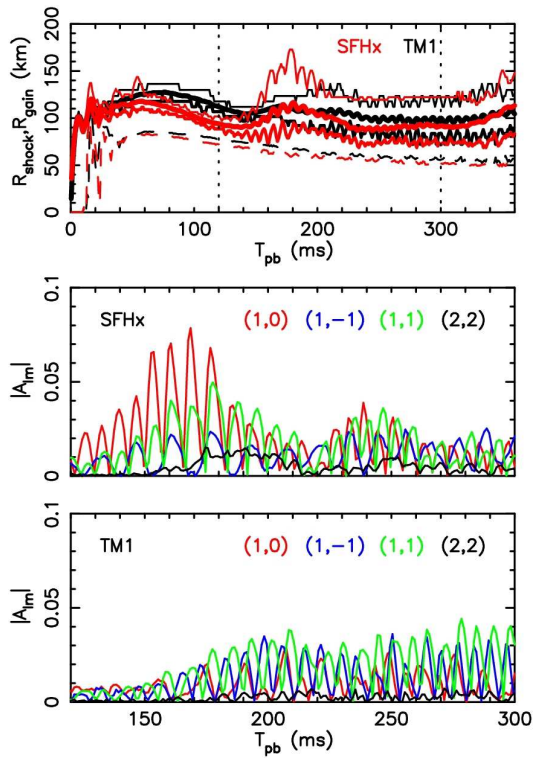}
\end{center}
  \caption{Top; Time evolution of maximum, average and minimum shock radii (solid)
  and spherically averaged gain radius (dashed) for models SFHx(red) and TM1(black).
  Two vertical dotted lines represent the period when the low frequency component ``B'' appears (Fig. \ref{GWs}). 
  Lower twos; Time evolution of normalized mode amplitudes $|A_{lm}|$ for several representative modes $(l,m)$ of SFHx(upper) and TM1(lower).
  We show the period bounded by two vertical dotted lines in top panel.
\label{Rshock.eps}}
\end{figure}

In Fig. \ref{Rshock.eps}, we plot time evolutions of maximum, average, minimum shock radii $R_{\rm shock}$ (top, solid)
and normalized mode amplitudes $|A_{lm}|\equiv |c_{lm}|/|c_{00}|$ (see \cite{Burrows12} for $c_{lm}$)
of spherical polar expansion of the shock surface $R_{\rm shock}(\theta,\phi)$.
For $A_{lm}$, we plot models SFHx(middle) and TM1(bottom) with focusing a period of $120\le T_{\rm pb}\le300$ ms which corresponds to
the appearance of component ``B''.
We also plot spherically averaged gain radius $R_{\rm gain}$ (dashed) in top panel.

The characteristic SASI motions seen in Fig.\ref{3Dimages} are reflected in the evolution of $|A_{lm}|$.
For SFHx, the most dominant mode during the first phase of the SASI (50 ms $\la T_{\rm pb}\la150$ ms)
is the sloshing mode, i.e. $(l,m)=(1,0)$, which is in accord with
the clear one sided shock heated region (top-left in Fig.\ref{3Dimages}).
Regarding the EoS dependence, although we do not see any qualitative differences between stiffest EoS model TM1
and the softest EoS one SFHx, TM1 shows less SASI development, i.e., smaller values of $|A_{lm}|$, during the SASI development phase.
DD2 also shows less SASI development compared to SFHx.
Such a quantitative difference can be explained by the shock radius.
In top panel of Fig. \ref{Rshock.eps}, TM1 shows more extended shock radii till $T_{\rm pb}\sim150$ ms.
This is because, depending on the stiffness of nuclear EoS, the bounce shock can be formed at larger radius
which can sometime amount to $\sim 0.1M_\odot$ difference in mass coordinate \citep{suwa13,Fischer14}.
Consequently the prompt shock has to plunge into more material and stalls at smaller radius in our softest EoS model SFHx.
The smaller shock radius is a favorable condition for the SASI development
due to the shorter advective-acoustic cycle \citep{Foglizzo02,Scheck08}.
Initial SASI activities reach their maxima when the shock expansion occurs due to sudden drop
of mass accretion rate at $T_{\rm pb}\sim150$ ms.
Afterward the spiral mode becomes dominant as seen in $A_{1\pm1}$ (see also top-right in Fig.\ref{3Dimages})
which lasts another $\sim150/200$ ms (SFHx/TM1).

In the final phase, the core experiences neutrino-driven convection
till the end of our calculation time $T_{\rm pb}\sim350$ ms.
During this phase, matters in the gain region are exposed intensively to neutrino radiations and form high entropy
$(s_{k_{\rm B}}\sim20)$ smaller scale convection plumes (bottom twos in Fig.\ref{3Dimages}).
Following \cite{Foglizzo06}, we check the parameter $\chi$.
Although $\chi\ga3$ is expected to be satisfied for convection to develop,
we find that $\chi$ stays $\sim0.5$ till $T_{\rm pb}\la350$ ms in both models
despite the appearance of convection plumes.
As already pointed out in \cite{Ott13,Hanke13}, this is because the initial perturbations in the gain region
are already not small when the neutrino convection phase initiates.
The gain radius ($R_{\rm gain}$ in Fig.\ref{Rshock.eps}) appears more inward
in SFHx which leads to higher entropic convection plumes compared to those in TM1
(compare bottom two panels in Fig. \ref{3Dimages}).


\begin{figure}[htpb]
\begin{center}
\includegraphics[width=80mm,angle=-90.]{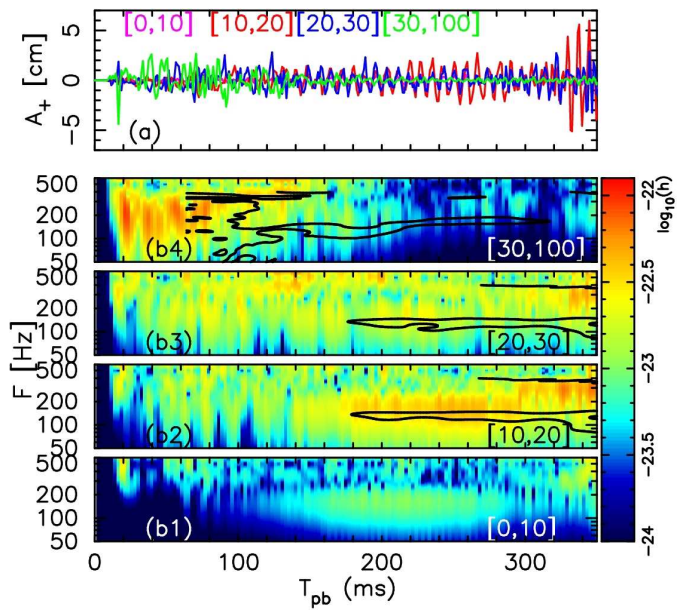}  
\end{center}
  \caption{Rough measurement of contribution from each spherical shell to (a) the GW amplitude and
  (b1-4) their spectrogram $\tilde h$ in a logarithmic scale. We show the contributions from four spherical shells
  with interval of [0,10], [10,20], [20,30] and [30,100] km.
  Black contours overplotted on spectrograms for $\tilde h$ represent half maximum of spectrograms for mass accretion rate
  measured at $R=17$(b2), 23(b3), and 48(b4) km.
\label{GwaveRadial}}
\end{figure}
Now, we discuss how these hydrodynamical evolutions affect on the GW emission ``B'' in Fig. \ref{GWs}.
By spatially decomposing the quadrupole moment of matters into several spherical shells,
we roughly localize this emission at $10\la R\la20$ km (Fig.\ref{GwaveRadial}).

Before going to further discussion, we present a back-of-the-envelope estimation of the GW amplitude as
{\setlength\arraycolsep{1pt}
\begin{eqnarray}
D|h|&\sim& 2\epsilon MR^2/T_{\rm dyn}^2\sim 2\epsilon M^2/R\sim 2\epsilon R^2\dot M^2/M,
\label{eq:Dh}
\end{eqnarray}}
where $M$, $R$ and $T_{\rm dyn}$ represent the mass, size and dynamical time scale of the system, respectively, in geometrized unit.
Here we have used the following reasonable assumptions
\begin{eqnarray}
\label{eq:Tdyn1}
T_{\rm dyn}\sim M/\dot M
\end{eqnarray}
or
\begin{eqnarray}
\label{eq:Tdyn2}
T_{\rm dyn}\sim R/V\sim\sqrt{R^3/M},
\end{eqnarray}
with $V\sim \sqrt{M/R}$ being the velocity derived by the energy conservation.
From the last relation in Eq.(\ref{eq:Dh}), we expect that significant time variation in the mass accumulation onto the PNS
can potentially lead to the GW emission.
In Fig.\ref{GwaveRadial}, we superimpose spectrogram of the mass accretion rate ${\dot M}(R)$
(the black contour at half maximum) measured at $R=17$, 23, and 48 km on top of the GW spectrogram.
While ${\dot M}(R=48{\rm km})$ starts quasi-periodic oscillation at $F\sim100-200$Hz around $T_{\rm pb}\sim120$ ms,
we find a time delay of $\sim60$ ms for their appearance at deeper region ($R=17$ and 23 km).
Since the density averaged mean radial velocity between the lepton driven ($10\la R\la 20$ km)
and the entropy driven ($R\ga 40$ km) convection layers is $\sim5\times10^7$ cm s$^{-1}$, the time delay is consistent
with the advection time scale over the stable layer ($20\la R\la 40$ km).
Furthermore, coincidence of time modulation in ${\dot M}(R)$ and the GW component ``B'' is obvious from panel (b2).

\begin{figure*}[htpb]
\begin{center}
\includegraphics[width=65mm,angle=-90.]{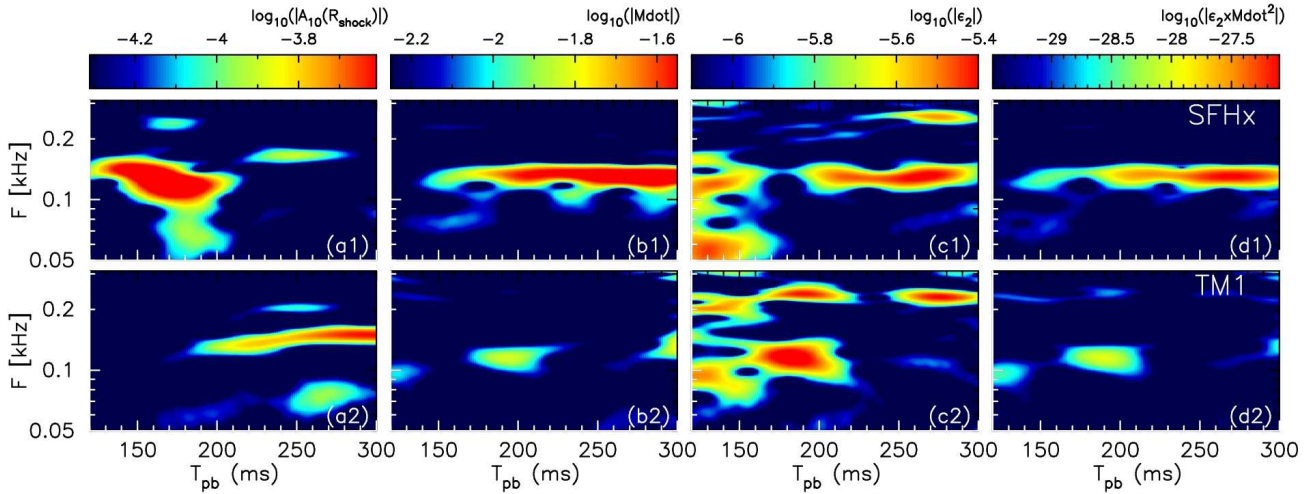}       
  \caption{Spectrograms of; (a) Fourier decomposed normalized mode amplitude $|\tilde A_{10}|$ of the shock surface
  for the sloshing-SASI mode, (b) the mass accretion rate $\tilde {\dot M}$ (with a dimension of $M_\odot$), through surface of a sphere
  with radius of $R=20$ km, (c) deformation of the isodensity surface $\tilde \epsilon_{l}$ for $l=2$ mode and
  (d) a rough measurement of the GW energy spectrum which is proportional to $\sim\epsilon R^2 {\dot{M}}^2M^{-1}$ (see text).
  Top and bottom rows are for SFHx and TM1, respectively.
  \label{SASIfrequency}}
\end{center}
\end{figure*}
Finally, to connect the SASI activities with the GW component B,
we plot spectrograms of normalized mode amplitude of the sloshing-SASI mode $|\tilde A_{10}|$,
the mass accretion rate $|\tilde {\dot M}|$ measured at $R=17$ km, 
normalised quadrupole deformation of the isodensity surface $\tilde \epsilon_{l}$ for $l=2$,
and a rough measurement of the GW energy spectrum in Fig. \ref{SASIfrequency}.
$\tilde \epsilon_{l}$ denotes a Fourier component of normalised mode amplitude $\epsilon_l$ defined by
\begin{eqnarray}
\label{eq:epsilon}
\epsilon_{l}\equiv \sqrt{\sum_{m=-l,l} \left( R^{14}_{l,m}\right)^2}\left/ R^{14}_{0,0}\right.,
\end{eqnarray}
where $R^{14}_{l,m}$ is evaluated by the spherical polar expansion of the isodensity surface $R^{14}$ extracted at $\rho=10^{14}$ g cm$^{-1}$
as the same way as for the shock surface.
Although several other modes are excited at the surface,
only the leading contribution ($l = 2$ mode) to the GW emission is shown in the panel.
As a reference, the isodensity surface $R^{14}$ locates $\sim13.5$ km during $150\la T_{\rm pb}\la300$ ms in SFHx.
From the last relation in Eq.(\ref{eq:Dh}), we plot $\log_{10}{|h|}\sim \log_{10}{\epsilon \dot M^2}+\rm{const.}$
in panels (d) of Fig. \ref{SASIfrequency} with assuming $M=0.5M_\odot$, a mass contained in $10\la R\la20$ km, and $R^{14}=13.5$ km
stay nearly constant.

During $140\la T_{\rm pb}\la180$ ms in SFHx, we see a strong sloshing motion which has its peak frequency
at $100\la F\la200$ Hz (a1).
With some time delay ($\sim50$ ms) from the appearance of it,
the mass accretion rate ${\dot M}$ starts showing a quasi-periodic oscillation at the same frequency range
$100\la F\la200$ Hz (b1) and it excites oscillation on the isodensity surface (c1).
A combination of large ${\dot M}$ and $\epsilon_2$ expect GW emissions appearing in panel (d1)
and it can well explain Fig. \ref{GWs}.
During $200\la T_{\rm pb}\la300$ ms, $\epsilon_{2}$ stays $\sim3\times10^{-4}$ in SFHx.
A rough measurement of the GW amplitude due to this deformation, $A\sim2\epsilon_2 M^2R^{-1}$,
deduces $A\sim2$ cm which is consistent with the actual amplitude (Fig.\ref{GwaveRadial}).

\section{Summary and Discussion}
We have presented relativistic 3D SN simulations with three different nuclear EoSs.
The overall pictures of SN dynamics are qualitatively the same among all three models,
although the development of the SASI differs quantitatively.
The softer the EoS is, the more the SASI develops, since the prompt shock stalls at smaller radii.
The evolution shows the first prompt convection phase, the sloshing-SASI phase
which shifts to the spiral mode and finally to the neutrino-driven convection phase.

Regarding the GWs, we have also confirmed 
previously reported emissions originated from
the PNS surface $g$-mode oscillation \citep{BMuller13,Murphy09}.
Additionally in the softest EoS model SFHx, in which the most vigorous SASI motion was observed,
we have found another low frequency ($100\la F\la200$Hz) quasi-periodic emission.
This emission was spatially localized at $10\la R\la20$ km.
Through a spectrogram analysis of the SASI modes,
of the mass accretion rate at $R=20$ km and of the quadrupole mode of the central core deformation,
we consider that the temporally perturbed mass accretion in association with the SASI downflows
penetrate into the PNS surface and excite the oscillation at $10\la R\la20$ km, which then leads to the GW emission.
Just recently, \cite{Andresen16} has also reported a similar result that the low frequency GW emission
occurs due to the SASI.
As another remarkable feature, these down flows also deform the neutrino spheres and cause time oscillation in the neutrino signals
\citep{Tamborra13}.
We will report the coherency between neutrinos and GW signals originated from the SASI activity in our upcoming paper.

At the end, we briefly discuss the detectability by the next generation of GW detectors, LIGO \citep{Harry10} and KAGRA \citep{Aso13}.
As for the PNS surface $g$-mode oscillation, we found a dependence on the nuclear EoS.
The peak frequency appears at $F=635$, 671, and 681 Hz in TM1, DD2, and SFHx, respectively,
which is in order of the stiffness of nuclear EoS.
At this frequency range, the signal-to-noise ratio ($S/N$), a simple comparison between the energy spectra and
sensitivity curves with assuming the source distance of $D=10$ kpc, reaches $\sim10$ for all the models.
Regarding the SASI-origin emission ``B'', which is observed only in SFHx,
the peak value of GW energy spectrum appears at $F=129$ Hz
and reaches almost a comparable amplitude to that from $g$-mode oscillation.
The $S/N$ reaches relatively high value of $\sim50$ 
because of that both detectors have their highest sensitivity at $\sim100-200$ Hz.
From these two spectral peak values, we expect that GWs from Galactic SNe,
even if their progenitors are non-rotating, are likely observable.
Following \cite{Hayama15}, we plan to perform a coherent
 network analysis for clarifying the detectability of these signals more in detail. 
 

\acknowledgements{
This work was supported by the European Research Council (ERC; FP7) under ERC
Advanced Grant Agreement N$^\circ$ 321263 - FISH.
TK acknowledges valuable comments and fruitful discussions with F.-K. Thielemann and M. Hempel.
Numerical computations were carried out on Cray XC30 at Center for Computational Astrophysics, National Astronomical Observatory of Japan.
This study was supported by JSPS (Nos.
24103006, 24244036, 26707013 and 26870823) and by
MEXT (Nos. 15H00789 and 15H01039) and
JICFuS as a priority issue to be tackled by using Post `K' Computer.
}




\end{document}